\documentstyle[twocolumn,aps,pra]{revtex}

\def\Db#1{{\langle}#1{|}}     
\def\Dsb#1{{\langle}#1}       
\def\Dk#1{{|}#1{\rangle}}     
\def\Dsk#1{#1{\rangle}}       

\begin{document}
\draft

\title{Schr\"{o}dinger revisited: an algebraic approach}
\author{M. R. Brown and B. J. Hiley}
\address{Theoretical Physics Research Unit, Birkbeck
College, University of London, Malet Street, London WC1E 7HX,
England}
\date{\today}

\maketitle
\begin{abstract}
Starting with the quantum Liouville equation, we write the density
operator as the product of elements respectively in the left and
right ideals of an operator algebra and find that the
Schr\"{o}dinger picture may be expressed through two representation
independent algebraic forms in terms of the density and phase
operators.  These forms are respectively the continuity equation,
which involves the commutator of the Hamiltonian with the density
operator, and an equation for the time development of the phase
operator that involves the anti-commutator of the Hamiltonian with
this density operator. We show that this latter equation plays two
important roles: (i) it expresses the conservation of energy in a
system where energy is well defined and (ii) it provides a simple
way to evaluate the gauge changes that occur in the Aharonov-Bohm,
the Aharonov-Casher, and Berry phase effects.  Both these operator
(i.e. purely algebraic)  equations also allow us to re-examine the
Bohm interpretation, showing that it is in fact possible to
construct Bohm interpretations in representations other than the
$x$-representation.  We discuss the meaning of the Bohm
interpretation in the light of these new results in terms of
non-commutative structures and this enables us to clarify its
relation to standard quantum mechanics.
\end{abstract}


\section{Introduction}
\label{sec:intro}

In a previous paper, Monk and Hiley \cite{MH} have suggested that
instead of using the traditional Hilbert space description of
quantum phenomena, one should give primary consideration to the
algebraic structure, not only because it has a number of
mathematical advantages that have already been pointed out by Dirac
\cite{PD45,PD65,PD66}, but because it offers the possibility of a
radically different interpretation of the quantum formalism. By
algebraic structures, we mean exploiting the rich possibilities
contained in, for example,  C* and W* (von Neumann) algebras, which
play an important role in field theory \cite{HK,RH} as well as in
equilibrium \cite{BR,JS} and non-equilibrium statistical mechanics
\cite {IP,GP78}.  In spite of the potential richness of these
methods, they have not been used in the general debate on the
foundations of quantum theory mainly because of the abstract nature
of the mathematics.

In their paper, Monk and Hiley \cite{MH} outlined how this
mathematics can be simplified so that the approach becomes much
more transparent.  Furthermore, one can develop an interpretation
for this formalism provided one is willing to give up the basic
ideas of particles- and/or fields-in-interaction and instead, think
in terms of process. Indeed, recent work \cite{NM,HM,MR95,MR98} has
shown how the algebraic approach does have a potential for taking
the discussions of the meaning of the quantum formalisms into new
domains. It is  this background that provides the motivation for
the present paper.  However,  we will not assume any detailed prior
knowledge of this abstract algebraic structure. Our purpose here is
to show how we can re-write the equations of elementary quantum
mechanics in a purely algebraic way, which, as we show, yields some
interesting new insights.

Traditionally, the algebraic approach has implied the use of the
Heisenberg picture.  Here the operators (or elements of the
algebra, in our case) become time dependent and carry the dynamics
of the quantum system. One clear advantage of this approach is that
all the elements of the algebra are representation-independent  and
so we are not tied to any one particular representation. A further
advantage is that the equations of motion  have a close structural
similarity to the classical equations of motion, viz,  commutator
brackets directly replace Poisson brackets \cite{PD45,PD65}.

In contrast to this, the Schr\"{o}dinger picture has the time
development entirely tied into the wave function, and as such is a
representation dependent object which  appears to exists only in
Hilbert space and does not appear in the algebra.   Thus, the
Schr\"{o}dinger picture does not seem to have the generality of the
representation-independent Heisenberg picture.

In this paper, we will show that this representation dependence of
the Schr\"{o}dinger equation  is only apparent and  in
section~\ref{sec:algse} we will show that it can be written in a
representation independent form.  This means writing the wave
function as a ``wave operator'' (in the left ideal of the algebra)
so that the Schr\"{o}dinger equation becomes purely algebraic and
independent of any representation in a Hilbert space.

The way to do this has been known for a long time \cite{RH,BR1}.
Indeed, Monk and Hiley \cite{MH} have already shown in simple terms
that the key step involves expressing the wave operator in the left
ideal. This means that the wave function must be replaced by the
density operator, $\rho$, even in the case of pure states. Here the
density operator plays the role of an idempotent and it is this
idempotency that is central in the whole approach.

Starting with the quantum Liouville equation and writing the
density operator as the product of elements respectively in the
left and right ideals in the algebra, we find that the
Schr\"{o}dinger picture may be expressed through two algebraic
forms which are representation independent. These forms are the two
equations~(\ref{eq:sec2e6}) and~(\ref{eq:sec2e7}) for the density
and phase operators. The first, expressed in terms of the
commutator of the Hamiltonian and the density operator, is an
alternative form of the quantum Liouville equation which, as is
well known, describes the conservation of probability.  The second
describes the time dependence of the phase and is expressed in
terms of the anti-commutator of the Hamiltonian and the density
operator. This second equation, which does not appear in the
literature as far as we are aware, becomes an equation for the
conservation of energy in systems when energy is well defined.

In section~\ref{sec:tqpe}, the time dependent phase operator
equation~(\ref{eq:sec2e7}) is shown to be gauge invariant and
reproduces some well known results of gauge theory  in a very
direct and simple way.  For example, one can immediately derive the
Aharonov-Bohm phase for a particle travelling in a vector potential
while a trivial extension incorporates the Aharonov-Casher phase.
This latter phase  arises when a neutral particle with a magnetic
moment passes a line-charge. In addition to these examples, the
Berry phase and its associated energy follow almost trivially from
the same equation.

In sections~\ref{sec:pc} and~\ref{sec:reba} we use these equations
to explore in more detail the Bohm interpretation [BI]
\cite{BH87,BH93}. Because the equations~(\ref{eq:sec2e6})
and~(\ref{eq:sec2e7}) are representation independent, we can
construct a BI based on trajectories in {\it any} representation.
To do this, we introduce a generalisation of the current density
operator and demonstrate its use in both ordinary space and
momentum space.  In section~\ref{sec:reba} and through the examples
in section~\ref{sec:spex} we show in detail how one can construct a
consistent BI in the $p$-representation. This is contrary to the
assertion that this is not possible even in ``the simplest case to
construct an acceptable causal interpretation'' \cite{DB53} and we
discuss the significance of this statement in the light of our
examples.

Our examples  not only  remove one of serious criticisms of this
interpretation, namely that it does not use the full symplectic
symmetry of the quantum formalism, but also provide us with new
insights into the meaning of BI and its relation to standard
quantum mechanics.  In all of this work no appeal is made to any
classical formalism whatsoever, showing that the BI is quantum
through-and-through.

Perhaps the most important conclusion of this work is to show the
BI arises directly from the non-commutative structure of the
quantum mechanical phase space.  Non-commutative geometries are not
built on any form of well defined continuous manifolds. We are
forced to {\it construct}  ``shadow manifolds'' \cite{HS97,HS98}.
As we show in section~\ref{sec:reba}, these shadow manifolds have
the structure of a phase space. One is constructed using the
$x$-representation and the other uses the $p$-representation. These
spaces are different but converge to the same  phase space in the
classical limit.

In the final section~\ref{sec:conc} we discuss the consequences of
this construction for the BI. In fact we show that our approach
more clearly illustrates the ideas that Bohm and Hiley presented in
the final chapter of their book \cite{BH93}. There it was argued
that a new way of exploring the meaning of the quantum formalism
required a new order, the implicate order,  this order having its
origins in the mathematics of non-commutative geometry \cite{DB80}.
The shadow phase spaces are examples of explicate orders. Finally,
we briefly discuss how the BI fits into this general scheme.

\section{The algebraic approach to the Schr\"{o}dinger equation}
\label{sec:algse}

We begin by writing the Schr\"{o}dinger equation in a general
representation
\begin{equation}
i\frac{\partial \psi (a_{i} ,t)}{\partial t} =H\psi (a_{i} ,t)\quad
\quad
\quad (\hbar =1)
\label{eq:sec2e1}
\end{equation}
where the $a_{i}$ are the eigenvalues of $A$ an algebraic element
or operator in the algebra\footnote{Throughout this paper, we set
$\hbar = 1$ and use the convention of representing operators by
capitals and eigenvalues by lower case letters.}.
Equation~(\ref{eq:sec2e1}) introduces the state vectors $\psi
({a}_{i}, t)$ that are not elements of the algebra; rather, they
are elements of a separate vector (Hilbert) space and as such
depend on a representation. In this form, the Schr\"{o}dinger
equation does not appear to be part of the algebra even though it
uses elements of the operator algebra.

On the other hand, in the Heisenberg approach we write the
Hamiltonian flow of the operator $A \in$ A as
\begin{equation}
A(0) \rightarrow A(t)= M(t)^{-1} A(0) M(t)
\end{equation}
where $M(t) = \exp[{-iHt}]$, so giving rise to the Heisenberg
equation of motion
\begin{equation}
\frac{dA}{dt} =\frac{1}{i}\left[ A,H\right]_{-}.
\end{equation}
This means that the Heisenberg time evolution can be regarded as an
inner automorphism in the algebra $M(t): A \rightarrow A \; \forall
\; A \in$ A.  The equation of motion can be generalised to include the explicit
time dependence of $A$ giving
\begin{equation}
\frac{dA}{dt} =\frac{\partial A}{\partial t} + \frac{1}{i}\left[ A,H\right]_{-} .
\label{eq:sec2e2}
\end{equation}
Note that this equation is representation-free and the time
evolution is discussed entirely within the algebra itself.

In the algebraic approach~\cite{BR1}, the state function is
introduced through a density operator, $\rho$. This operator is
actually in the quantum algebra so that the Schr\"{o}dinger time
development must be implicit within the algebra itself. How, then,
can the time evolution be \textit{algebraically} expressed within
the Schr\"{o}dinger picture, without reference to either Hilbert
space or a particular representation in it?

In the usual approach to quantum mechanics, the density operator
is, unfortunately, not introduced as a primitive notion in the
theory. Rather, it is introduced almost as an after-thought when it
is found necessary to deal with mixed states. But using the density
operator as a starting point has the advantage of including both
pure states and mixed states together and of satisfying the
idempotent condition $\rho = {\rho}^2$.   Moreover, if we adopt the
further defining condition that the density operator must satisfy
the Liouville theorem
\begin{equation}
\frac{d \rho}{d t} = 0,
\end{equation}
then, since the density operator is also an element of the algebra,
equation~(\ref{eq:sec2e2}) immediately leads to the quantum
Liouville equation
\begin{equation}
\frac{\partial \rho}{\partial t} = \frac{1}{i}\left[ H, \rho
\right]_{-}.
\label{eq:sec2e3}
\end{equation}

If $\rho = \Dsk{\psi}\Dsb{\psi}$, is an element of the algebra,
then it must also be possible to identify the ket and the bra with
particular elements of the algebra.  A ket is an element of a
vector space, which when multiplied from the left it must remain in
that space. The algebraic equivalent of this vector space is a left
ideal I$_{L}$. An element, $B \varepsilon$, of a left ideal, where
$B$ is a wave operator and $\varepsilon$ a primitive idempotent,
corresponds to a ket. Similarly an element, $\varepsilon C$, of a
right ideal, I$_{R}$, corresponds to a bra. This suggests we write
$\rho= B\varepsilon \varepsilon C = B
\varepsilon C$, which means that a pure state density operator
corresponds to a two sided ideal, subject to the conditions $\rho =
{\rho}^2$ and tr$\rho = 1$.

To see how to write the algebraic equivalent to the Schr\"{o}dinger
equation, let us substitute $\rho= B \varepsilon C$ into the
equation of motion~(\ref{eq:sec2e3}), so that
\begin{equation}
i\hbar\left( \frac{\partial B}{\partial t} \right) \varepsilon C +
i\hbar B \varepsilon \left(
\frac{\partial C}{\partial t} \right)= H B \varepsilon C - B \varepsilon C H.
\end{equation}
Since $B$ and $C$ are operator elements outside the ideals of the
algebra, it can be assumed that there exist $B^{\dag}
: B^{\dag}B = \mathbf{1}$ and $C^{\dag} : C C^{\dag} = \mathbf{1}$.
Multiplying the above equation from the left by ${B}^{\dag}$ and
from the right by ${C}^{\dag}$, we find after re-arrangement, that
\begin{equation}
B^{\dag} \left( i \hbar \frac{\partial B}{\partial t} - HB
\right) \varepsilon
=
-  \varepsilon \left( i \hbar \frac{\partial C}{\partial t} + CH \right) C^{\dag}.
\end{equation}
Since $B (C)$ is \textit{any} non-null element of the algebra, we
can write
\begin{equation}
i\left( \frac{\partial B}{\partial t} \right) \varepsilon = HB
\varepsilon
\label{eq:sec2e4}
\end{equation}
and
\begin{equation}
- i\varepsilon \left( \frac{\partial C}{\partial t} \right) = \varepsilon C H.
\label{eq:sec2e5}
\end{equation}
We see immediately that equations~(\ref{eq:sec2e4})
and~(\ref{eq:sec2e5}), which are respectively in the left and right
ideals of the algebra, have the same general form as the
Schr\"{o}dinger equation and its conjugate counterpart ($H$ is
assumed to be Hermitian). We stress here again that $B$ and $C$ are
elements of the
\textit{algebra} and not elements of a Hilbert space.

To see exactly how these two equations are related to the usual
Hilbert space formalism, we specifically choose the wave operator
$B$ to be a function of the position operator $X$ so that $B
\varepsilon = B(X,t)
\in$ I$_{L}$ and then project $B(X,t)$ into a complex function
belonging to $L^{2}(x,\mu)$ viz,\footnote{$L^{2}(x,\mu)$ means
square integrable complex functions with measure $\mu$.}
\begin{equation}
\eta : B(X,t) \rightarrow B(x,t)
\end{equation}
so that
\begin{equation}
B(X,t)(x) = B(x,t).
\end{equation}
This is the usual wave function\footnote{Not all elements of a left
ideal produce state functions that are physically meaningful. We
will not discuss these restriction here. (See
Ballentine~\cite{BA90}.)}, conventionally written as $\Psi(x,t)$.
It is now straight forward to show that equation~(\ref{eq:sec2e4})
becomes the Schr\"{o}dinger equation
\begin{equation}
i\frac{\partial \Psi (x,t)}{\partial t} =H(x)\Psi (x,t).
\end{equation}
The conjugate equation can be derived by first assuming the dual
projection
\begin{equation}
{\eta}* : C(X,t) \rightarrow C^{*}(x,t)
\end{equation}
so that
\begin{equation}
C(X,t)(x)=C^{*}(x,t).
\end{equation}
Again it is straight forward to show that
equation~(\ref{eq:sec2e5}) leads to the conjugate Schr\"{o}dinger
equation. Thus equations~(\ref{eq:sec2e4}) and~(\ref{eq:sec2e5})
are the algebraic, representation independent equivalents of the
Schr\"{o}dinger equation.

Now let us continue developing the general structure.  We write the
wave operators $B$ and $C$ in the mutually conjugate forms $B =
\exp[i S_{Q}(t)]$ and $C= \exp[- i S_{Q}^{\dag}(t)]$ where $S_{Q}=
S - i \ln R$. In this case equations~(\ref{eq:sec2e4})
and~(\ref{eq:sec2e5}) become the dual pair
\begin{equation}
-\frac{\partial S_{Q}}{\partial t} B \varepsilon = H B \varepsilon
\label{eq:sec2e45a}
\end{equation}
and
\begin{equation}
- \varepsilon C \frac{\partial S_{Q}^{\dag}}{\partial t}  = \varepsilon C H,
\label{eq:sec2e45b}
\end{equation}
which are quantum algebraic
equivalents\footnote{Equations~(\ref{eq:sec2e45a})
and~(\ref{eq:sec2e45b}) cannot be simplified further, since $B
\varepsilon \in I_{L}$ and $\varepsilon C \in I_{R}$ do not have
inverses.} of the Hamilton-Jacobi equation of classical mechanics
\begin{equation}
\frac{\partial S_{cl}}{\partial t} + H = 0,
\end{equation}
where $S_{cl}$ is the classical action.  It is natural, therefore,
to call $S_{Q}$ the quantum action.

Equations~(\ref{eq:sec2e45a}) and~(\ref{eq:sec2e45b}) respectively
evolve in the left and right ideals, which are mutually dual
spaces, so reflecting the essential duality between the
Schr\"{o}dinger equation and its complex conjugate. This duality
can be lifted out of the left and right ideals of the algebra and
reflected in another pair of algebraic equations. Post- and
pre-multiplying equations~(\ref{eq:sec2e45a})
and~(\ref{eq:sec2e45b}) by $\varepsilon C$ and $B \varepsilon$
respectively and then adding and subtracting the resulting
equations, we find
\begin{equation}
\left( -\frac{\partial S_{Q} }{\partial t} \rho +\rho
\frac{\partial S^{\dag}
_{Q} }{\partial t} \right) =\left[ H, \rho\right] _{-}
\label{eq:sec2e6}
\end{equation}
and
\begin{equation}
\left( -\frac{\partial S_{Q} }{\partial t} \rho -\rho
\frac{\partial S^{\dag}
_{Q} }{\partial t} \right) =\left[ H, \rho\right] _{+}
\label{eq:sec2e7}
\end{equation}
Of note is the appearance of the commutator and the anti-commutator
with the Hamiltonian on the RHS of these two equations. They are
mathematically equivalent to the Schr\"{o}dinger equation and its
conjugate and are general in the sense that they are independent of
a specific representation. Expressing equation~(\ref{eq:sec2e6}) in
Hermitian form, we immediately recover~(\ref{eq:sec2e3}), the
quantum Liouville equation.  Equation~(\ref{eq:sec2e7}) cannot in
general be reduced to a simpler algebraic form but may be
recognised as a symmetrised operator form of the Hamilton-Jacobi
equation.

The meaning of equations~(\ref{eq:sec2e6}) and~(\ref{eq:sec2e7})
can be further clarified by looking at the diagonal elements in the
$a$-representation. Thus,
\begin{equation}
\frac{\partial {\rho}_{R}(a)}{\partial t} + \frac{1}{i}\left[ \rho ,H\right] _{-} (a)=0
\label{eq:sec2e8}
\end{equation}
and
\begin{equation}
{\rho}_{R}(a)\frac{\partial S(a)}{\partial t} +\frac{1}{2} \left[
\rho ,H\right]
_{+} (a)=0.
\label{eq:sec2e9}
\end{equation}
Here we have written [\dots .](a) = $\Db{a} \left[ ....\right]
\Dk{a}$ and ${\rho}_{R}=R^2$.  If we now choose $A$ to be the position operator, then
equation~(\ref{eq:sec2e8}) takes the form
\begin{equation}
\frac{\partial \mathcal{P}}{\partial t} + \nabla \cdot \bf{j} =0
\label{eq:sec2e10}
\end{equation}
where ${\mathcal{P}} = {\mathcal{P}}(\bf{x}) = {\rho}_{R}(\bf{x}) =
\Db{\bf{x}} \rho \Dk{\bf{x}}$ and $\bf{j}$ is a probability
current. Thus, the Liouville equation~(\ref{eq:sec2e8}) is identified
with the
\textit{conservation of probability} as expected.
Equation~(\ref{eq:sec2e9}) describes the time variation of the
quantum phase and so we will call it (and the more general
form~(\ref{eq:sec2e7})) the quantum phase equation. In a state in
which the energy is well defined, this equation becomes
\begin{equation}
\frac{\partial S}{\partial t} =-E.
\label{eq:sec2e11}
\end{equation}
In this case, equation~(\ref{eq:sec2e9}) expresses the
\textit{conservation of energy} in Hamilton-Jacobi form.

The Liouville equation~(\ref{eq:sec2e8}) is well known and plays a
prominent role in quantum statistical mechanics. The quantum phase
equation~(\ref{eq:sec2e9}) does not usually appear in the
literature, although something similar has been used by George et
al.~\cite{GP78} in their discussions of irreversible quantum
processes. In their case, the anti-commutator is simply introduced
by defining it to be the energy super-operator. What we show here
is that this operator comes directly from the Schr\"{o}dinger
equation and although the extension to super operator status is
possible, this generalisation is not necessary for the purposes of
this paper.

In summary then, equations~(\ref{eq:sec2e6}) and~(\ref{eq:sec2e7})
are simply the algebraic equivalents of the Schr\"{o}dinger
equation when it is written in a way that does not depend on a
specific representation. It is now easy to confirm that these two
equations, when expressed in a particular representation, are
simply the real and imaginary parts of the Schr\"{o}dinger equation
under a polar decomposition of the wave function written in that
particular representation.

\section{The quantum phase equation}
\label{sec:tqpe}

\subsection{Gauge invariance}

We will first examine equation~(\ref{eq:sec2e9}) in some detail.
Let us begin by looking at the form of this equation in the
$x$-representation when we choose the Hamiltonian  $H = p^{2}/2m +
V(x)$.  Here equation~(\ref{eq:sec2e9}) becomes
\begin{equation}
\frac{\partial S}{\partial t} +
\frac{1}{2m}\left( \frac{\partial S}{\partial x}\right)^{2} + V(x)
-\frac{1}{2mR}\left(\frac{\partial ^{2}R}{\partial x^{2}}\right ) = 0
\label{eq:n8}
\end{equation}
This equation is the real part of the Schr\"{o}dinger equation in
the $x$-representation.

Before  examining this equation in detail we must ensure that it is
gauge invariant.  To show that this is the case, let us first
introduce the  gauge transformation $V'(x) = V(x) + V_{0}$.  This
must be accompanied by the phase transformation  $\psi'(x, t) =
\phi (x) \exp[-i(E + V_{0})t]$.  Since we are considering a well
defined energy state, the transformed equation~(\ref{eq:sec2e7})
will read
\begin{equation}
((- \frac{\partial S_{Q}}{\partial t} + V_{0})\rho -
\rho (\frac{\partial S_{Q}^{\dag}}{\partial t} + V_{0}))
= [H,\rho]_{+} + [ V_{0},\rho]_{+}
\end{equation}
because $\rho ' = \rho$, $H ' = H + V_{0}$ and $S_{Q} '
= S_{Q} + V_{0}$.  Then, since $\left[V_{0},\rho \right]_{+}
= 2 V_{0} \rho$, we immediately recover equation~(\ref{eq:sec2e7}),
so establishing its gauge invariance and that of
equation~(\ref{eq:sec2e9}).

Gauge invariance in this case involves the phase change $S' = S + S_{0}$ then since
\begin{equation}
V'(x, t) = V(x) + V_{0}(t)
\end{equation}
our equation gives
\begin{equation}
\frac{\partial S_{0}}{\partial t} = V_{0}(t)
\end{equation}
so that
\begin{equation}
S_{0} = \int_{t_{0}}^{t} V_{0}(t')dt'
\end{equation}
It will immediately be recognised that this is the expression for
the scalar part of the Aharonov-Bohm effect \cite{AB}.

We can also obtain the magnetic phase shift from
equation~(\ref{eq:sec2e9}) by starting from the Hamiltonian
\begin{equation}
H = \frac{1}{2m}({\bf{P}}- e{\bf{A}})^{2}\hspace{2cm}(c = 1)
\end{equation}
On expanding this Hamiltonian, we find
\begin{eqnarray}
H &= \frac{1}{2m}{\bf{P}}^{2} - \frac{e}{2m}\left (\bf{P
\cdot A} +
\bf{A \cdot P} \right) + \frac{e^{2}}{2m}A^{2}\\ &= H_{
free particle} + H_{int} + H_{free field}
\end{eqnarray}
The corresponding phase will then consist of three terms
\begin{equation}
S = S_{free particle} + S_{int} + S_{free field}
\end{equation}
so that the diagonal form
\begin{equation}
\rho_{R} \left( \frac{\partial S_{int}}{\partial t}\right) +
\frac{1}{2} \left[ H_{int},\rho \right]_{+}= 0
\end{equation}
gives
\begin{equation}
\rho_{R}\frac{\partial S_{int}}{\partial t} = e
{\bf{A}} \cdot {\bf{j}}_{x}
\end{equation}
in the $x$-representation. If we then write ${\bf{j}}_{x} =
\rho_{R} {\bf{v}}$, we find
\begin{equation}
 S_{{int}} = e \int_{t_{0}}^{t} {\bf{A \cdot v}} dt = e
 \int_{x_{0}}^{x}{\bf{A}} \cdot d\bf{x}
\end{equation}
We immediately recognise this equation as the expression for the
Aharonov-Bohm phase for the vector potential\footnote{ This effect
was derived in the $x$-representation from the `guidance' condition
by Philippidis et al \cite{PBK}. A more recent discussion using
this approach rather than the method we use can be found in
Sj\"{o}qvist and Carlsen \cite{SC}}.

The phase for the Aharonov-Casher effect \cite{AC}, which involves
a neutral particle with a magnetic moment passing a line of
electric charges, also follows trivially once the Hamiltonian
\begin{equation}
H = \frac{1}{2m}\left({\bf{P}} - {\bf{E}} \times {\bf{\mu}}
\right)^{2} - \frac{{\bf{\mu}}E^{2}}{m}
\end{equation}
is assumed.  The additional phase change also
follows trivially from the same procedure used for the vector potential.

It should also be noted that the Berry phase \cite{MB,AA} emerges
directly from equation~(\ref{eq:sec2e9}).  In this case the
behaviour of the quantum system depends on some additional cyclic
parameter ${\cal B}(t)$. The phase now becomes a function of this
parameter. Thus equation~(\ref{eq:sec2e9}) becomes
\begin{equation}
\rho_{R} \left( \frac{\partial S}{\partial t} + \dot{{\cal B}}
\frac{\partial S}{\partial {\cal B}}\right)
+ \frac{1}{2}\left[\rho, H \right]_{+}  = 0
\end{equation}
giving an extra phase factor $\dot{{\cal B}} \frac{\partial
S}{\partial {\cal B}}$.  Thus the contribution to the phase from
this extra degree of freedom is
\begin{equation}
S_{{Berry phase}} = \int_{t_{0}}^{t}\dot{{\cal B}}
\frac{\partial S}{\partial {\cal B}}dt
\end{equation}
To  evaluate this term, we need to consider specific problems which
means going to a specific Hamiltonian in a specific representation.
This representation is generally the $x$-representation.  Berry
\cite{MB} considered the case of the precession of  nuclear spin in
a magnetic field in his original paper, and showed that
\begin{equation}
 \frac{\partial S}{\partial {\cal B}} = \Im \langle
 {\cal B}(t), t|\nabla_{{\cal B}}|{\cal B}(t), t\rangle
\end{equation}
so that
\begin{equation}
S_{{Berry phase}} = \Im \int_{{\cal B}_ {0}}^{{\cal B}}\langle {\cal
B}(t), t|\nabla_{{\cal B}}|{\cal B}(t), t\rangle d{\cal B}
\end{equation}
which is exactly the result obtained by Berry \cite{MB}.

\subsection{The $x$- and $p$-representations}

Having seen how the  additional phase changes arise for these
simple gauge fields, we now return to examine the details of
equation~(\ref{eq:n8}). To do this, let us consider the case of the
harmonic oscillator with Hamiltonian $H = p^{2}/2m + Kx^{2}/2$. In
this case equation~(\ref{eq:n8}) reads
\begin{equation}
\frac{\partial S_{x}}{\partial t} + \frac{1}{2m}
\left( \frac{\partial S_{x}}{\partial x}\right)^{2} +
\frac{Kx^{2}}{2} -\frac{1}{2mR_{x}}
\left(\frac{\partial ^{2}R_{x}}{\partial x^{2}}\right ) = 0
\label{eq:n12}
\end{equation}
where we have inserted the suffix $x$ to emphasise that this is
equation~(\ref{eq:sec2e9}) expressed in the $x$-representation.

Now let us write down the
corresponding equation in the $p$-representation. This takes the form
\begin{equation}
\frac{\partial S_{p}}{\partial t}+\frac{p^{2}}{2m}+\frac{K}{2}
\left(\frac{\partial
S_{p}}{\partial p}\right)^{2}-\frac{K}{2R_{p}}\left(\frac{\partial
^{2}R_{p}}{\partial p^{2}}\right)=0
\label{eq:n13}
\end{equation}

It should be noted that although the functional forms of these two
equations are clearly different, they nevertheless have the same
energy content.  This can be very easily checked for the ground
state of the harmonic oscillator.  One can quickly show that  both
equations give the  ground state energy to be $\omega
/2$, the zero-point energy.

In spite of the differences in functional form, there are
structural similarities between these two equations. These arise
essentially because we have chosen a symmetric Hamiltonian.  For
example, instead of the $p$ that appears in what looks like a
kinetic energy term in  equation~(\ref{eq:n13}), we have $(\partial
S_{x}/\partial x)$ in equation~(\ref{eq:n12}), and instead of $x$
in the potential energy term in equation~(\ref{eq:n13}), we have
$(\partial S_{p}/\partial p)$.  The last term in each equation has
the same general form except with the roles of $x$ and $p$
interchanged.

Since equation~(\ref{eq:n12}) is the real part of the
Schr\"{o}dinger equation, we can identify\footnote{Holland
\cite{PRH} has called expressions of this type `local expectation
values'.}
\begin{equation}
p_{r} =  \frac{\Re \left[\psi^{*}(x)P\psi(x)\right]}{|\psi(x)|^{2}}
= \left(\frac{\partial S_{x}}{\partial x}\right)
\label{eq:n14}
\end{equation}
Substituting this into equation~(\ref{eq:n12}) we find
\begin{equation}
\frac{\partial S_{x}}{\partial
t}+\frac{p_{r}^{2}}{2m}+\frac{K}{2}x^{2}-\frac{1}{2mR_{x}}
\left(\frac{\partial
^{2}R_{x}}{\partial x^{2}}\right)=0
\label{eq:n15}
\end{equation}
In the Bohm interpretation, $p_{r}$ was identified with the
``beable'' momentum. With this identification
equation~(\ref{eq:n14}) makes it now quite clear why the beable
momentum is a function of $x$, in contrast to the classical
momentum which is always an independent variable.

If $p_{r}$ is a momentum, then clearly equation~(\ref{eq:n15})
looks like an equation for the total energy of the quantum system.
If we make the assumption that this is an expression for the
conservation of energy then, in quantum theory, we must have an
additional quality of energy represented by the last term on the
RHS. This term is, of course, the quantum potential energy.  As has
been shown elsewhere this new quality of energy offers an
explanation of quantum processes like interference, barrier
penetration and quantum non-separability, all of which are quantum
phenomena \cite{BH93}.

Notice that equation~(\ref{eq:n14}) allows the possibility of
approaching the classical limit smoothly.   In this limit $S_{x}
\rightarrow S_{cl}$,  $p_{r} \rightarrow p_{cl} = (\partial
S_{cl}/\partial x)$ and the quantum potential energy becomes
negligible so that equation~(\ref{eq:n15}) becomes the classical
Hamilton-Jacobi equation.

Before continuing, we wish to stress a point that has not been
often fully appreciated, namely, that equation~(\ref{eq:n12}) is a
quantum equation and converting it to equation~(\ref{eq:n15})
requires no appeal whatsoever to classical physics. It is true that
in the traditional approach to this equation, the BI has made use
of the relation $p = (\partial S_{x}/\partial x)$ by appealing to
classical canonical theory.  However, this is an unnecessary
backward step.

It is because equations~(\ref{eq:sec2e9}) and~(\ref{eq:n12}) are
part of the quantum formalism that we were able to derive quantum
effects such as the Aharonov-Bohm, Aharonov-Casher and Berry phases
from equation~(\ref{eq:sec2e9}). In passing it should also be noted
that both the $x$- and $p$-representations of this equation (i.e.,
~(\ref{eq:n12}) and~(\ref{eq:n13}) ) contain a term which we have
called the quantum potential. This potential is modified by the
presence of the gauge effects as was first shown by  Philippidis,
Bohm and Kaye \cite {PBK}.   The quantum potential is central to
ensuring energy is conserved and, furthermore, it encapsulates
quantum non-separability or quantum non-locality~\cite{BH75}. The
quantum potential plays a key role in our approach and must be
distinguished from Bohmian mechanics as advocated by D\"{u}rr et
al.\cite{DG}.

If we now turn to the $p$-representation, i.e.,
equation~(\ref{eq:n13}),  we can write it in the form
\begin{equation}
\frac{\partial S_{p}}{\partial t} + \frac {p^{2}}{2m} +
\frac{K}{2}x_{r}^{2} - \frac{K}{2R_{p}}
\left(\frac{\partial ^{2}R_{p}}{\partial
p^{2}}\right) = 0
\label{eq:n16}
\end{equation}
by introducing
\begin{equation}
x_{r} = \frac{\Re\left[\psi^{*}(p)X\psi(p)\right]}{|\psi(p)|^{2}}
= - \left(\frac{\partial S_{p}}{\partial p}\right)
\label{eq:n17}
\end{equation}
Here $x_{r}$ is the position ``beable'', which now supplements the
momentum $p$.  Again in the classical limit,  we have $S_{p}
\rightarrow S_{cl}$, $x_{r}
\rightarrow x_{cl} = - (\partial S_{cl}/\partial p)$ and the
last term on the RHS
of~(\ref{eq:n16}) becomes negligible. It should be noted that in
this limit equations~(\ref{eq:n15}) and~(\ref{eq:n16}) reduce to
the same equation giving rise to a unique phase space, which is
identical to the classical phase space.

All the above equations are part of standard quantum mechanics.
Although we have drawn attention to the significance of
equations~(\ref{eq:n15}) and~(\ref{eq:n16}) to the BI, we have yet
to discuss the interpretation in any detail.  To do this we first
need to find a way to calculate ``trajectories''.  In the
traditional approach to the BI this is done by regarding  $p =
(\partial S_{x}/\partial x)$ as a ``guidance'' condition and then
using $\dot{x}= p/m$ from
 which one can calculate a set of trajectories.  These trajectories
are then integrals of the velocity associated with the probability
current in the co-ordinate representation. However this is not the
general way to do it as can be seen by considering the
$p$-representation. The analogous expression in this representation
is $ x = -(\partial S_{p}/\partial p)$ and this clearly cannot be
regarded as a ``guidance'' condition.  Something is not quite right
here.  In order to find out what is involved it is necessary to
explore the Liouville equation~(\ref{eq:sec2e8}) in more detail.

\section{ Probability currents}
\label{sec:pc}

In this section we will focus our attention on the Liouville
equation~(\ref{eq:sec2e8}).  In the $x$-representation, this
equation gives rise to the conservation of probability
equation~(\ref{eq:sec2e10}) with the probability current defined by
\begin{equation}
{\bf{j}}=\frac{1}{2mi}[\psi^{*}(\nabla \psi) - (\nabla \psi
^{*})\psi].
\label{eq:n19}
\end{equation}
However, our aim is to find an expression for the current that is
not representation specific.  To do this we first consider the
classical Liouville equation
\begin{equation}
\frac{\partial \rho}{\partial t} + \{ \rho, H \} = 0
\end{equation}
where $\{É\}$ is the Poisson bracket. It is easy to verify that
this equation can be written in the form
\begin{equation}
\frac{\partial \rho }{\partial t} +
\{ {\bf{j}}_{x}^{c}, {\bf{p}} \} -
\{ {\bf{j}}_{p}^{c}, \bf{x} \} = 0
\label{eq:n20}
\end{equation}
with
\begin{equation}
{\bf{j}}_{x}^{c}= \rho \nabla_{p} H
\hspace{1cm}\makebox{and} \hspace{1cm} {\bf{j}}_{p}^{c} = -
\rho\nabla_{x} H.
\label{eq:n21}
\end{equation}
(The Poisson bracket of two vector functions is defined here as $\{
{\bf{v}},{\bf{w}} \} = {\sum}_{k} \{ v_k , w_k \}$.) If we now follow
Dirac's suggestion by respectively replacing classical variables and
Poisson brackets with operators and commutators, we find
\begin{equation}
 i\frac{\partial \rho}{\partial t} +[{\bf{J}}_{X}, {\bf{P}}] -
 [{\bf{J}}_{P},{\bf{X}}] = 0
\label{eq:n22}
\end{equation}
with
\begin{equation}
{\bf{J}}_{X} = \nabla_{P} (\rho H) \hspace{1cm}\makebox{and}
\hspace{1cm}{\bf{J}}_{P} = -
\nabla_{X} (\rho H)
\label{eq:n23}
\end{equation}
where the derivatives are on operators.  For $f(\rho,X,P)
= \rho X^{k}P^{n} \; (k \ge 1, n \ge 1)$ these are defined as \cite{BJ25}
\begin{equation}
\frac{\partial f}{\partial X}  = X^{k-1}P^{n}\rho +
X^{k-2}P^{n}{\rho}X + {...} + P^{n}{\rho}X^{k-1}
\end{equation}
and
\begin{equation}
\frac{\partial f}{\partial P}  = P^{n-1}{\rho}X^{k} +
P^{n-2}{\rho}X^{k}P + {...} + {\rho}X^{k}P^{n-1}.
\end{equation}
In the simple case of a free particle of mass $m$ we have
\begin{equation}
{\bf{J}}_{X} = \frac{1}{2m} (\rho {\bf{P}} + {\bf{P}}
\rho) \;\; \textnormal{and} \;\; {\bf{J}}_{P}  =  0.
\end{equation}

To see how this connects to the conventional results, let us
evaluate equation~(\ref{eq:n22}) in the $x$-representation. Here we
find
\begin{equation}
i\frac{\partial \Db{{\bf{x}}} \rho \Dk{{\bf{x}}}}{\partial
 t} + \Db{{\bf{x}}}[{\bf{J}}_{X},{\bf{P}}]
\Dk{{\bf{x}}}
-\Db{{\bf{x}}}[{\bf{J}}_{P},{\bf{X}}]\Dk{{\bf{x}}} = 0.
\label{eq:n24}
\end{equation}
If $H = \frac{{\bf{P^{2}}}}{2m} +V({\bf{X}})$ then the first
commutator gives $\Db{{\bf{x}}}[{\bf{J_{X}}},{\bf{P}}]\Dk{{\bf{x}}}
= \nabla_{x} \cdot {\bf{j}}_{x}$  and the second commutator vanishes.
Thus, equation~(\ref{eq:n24}) becomes
\begin{equation}
\frac{\partial {\mathcal{P}}({\bf{x}})}{\partial t} +
\nabla_{x} \cdot {\bf{j}}_{x} = 0
\label{eq:n25}
\end{equation}
which is just equation~(\ref{eq:sec2e10}) and
\begin{equation}
{\bf{j}}_{x} = \Db{{\bf{x}}}\nabla_{P}(\rho
\frac{{\bf{P^{2}}}}{2m})\Dk{{\bf{x}}}
\label{eq:n26}
\end{equation}
which is gives an  expression for the current that is identical to
the usual expression given by equation~(\ref{eq:n19}). Furthermore,
it is unique since it is independent of the form of the potential
used in the Hamiltonian.

In the $p$-representation we find
\begin{equation}
\frac{\partial {\mathcal{P}}({\bf{p}})}{\partial t} +
\nabla_{p} \cdot {\bf{j}}_{p} = 0
\label{eq:n27}
\end{equation}
where
\begin{equation}
{\bf{j}}_{p} = - \Db{{\bf{p}}}\nabla_{X}(\rho
V({\bf{X}}))\Dk{{\bf{p}}}
\label{eq:n28}
\end{equation}
Thus, we can now calculate probability currents in the
$p$-representation.  Unfortunately equation~(\ref{eq:n28}) does not
give us a model independent expression for the probability current
because the specific form of the current  depends on the form of
$V(x)$.  On reflection this is not surprising because the rate of
change of momentum must depend upon the externally applied
potential.  We will examine the consequences of these results for
the Bohm interpretation in the next section.

\section{Re-examination of the Bohm approach}
\label{sec:reba}

Let us now re-appraise the Bohm approach in the light of the new
results presented above.

It has been assumed that it is not possible to construct a BI using
any representation other than the $x$-representation.  This belief
arises from an early correspondence between Epstein \cite{SE} and
Bohm \cite{DB53}. Epstein  suggested that it should be possible to
develop an alternative causal interpretation by starting in the
momentum representation.  Bohm replied agreeing that a new causal
interpretation could possibly arise  from such a procedure provided
the canonical transformation on the particle variables were
simultaneously accompanied by a corresponding linear transformation
on the wave function. But, he concluded that this did not seem to
lead, even in the simplest of cases,  to an acceptable causal
interpretation.  He does not explain why he came to this conclusion
but this position has remained the accepted wisdom.

The general results with the harmonic oscillator presented above
show that, at least as far as the mathematics is concerned, it does
seem possible to develop a causal interpretation in the
$p$-representation based upon
equations~(\ref{eq:n16}),~(\ref{eq:n17}) and~(\ref{eq:n28}). We
will illustrate how this can be done using specific examples in
section~\ref{sec:spex}, but here we will simply discuss the general
principles involved.

We have already pointed out in section~\ref{sec:tqpe} that the so
called ``guidance'' condition is assumed to play a pivotal role in
what is known as ``Bohmian mechanics'' \cite{DG} does not
generalise to the $p$-representation.  However what does generalise
is a method based on probability currents.  Thus, in any
$q$-representation we use
\begin{equation}
\frac{dq}{dt} = \frac{j_{q}}{{\mathcal{P}}(q)}
\end{equation}
which can be integrated immediately to find a set of trajectories
in a general $(q,t)$ space\footnote{Hereafter, the notation is
restricted to one degree of freedom. Generalisation to many degrees
of freedom is straight forward.}.

In the $x$-representation we have
\begin{equation}
\frac{dx}{dt} = \frac{j_{x}}{{\mathcal{P}}(x)}
\end{equation}
which, when integrated, gives the particle trajectories of the BI.
This approach is similar to that used in pragmatic quantum
mechanics where the probability current is assumed to describe the
flux of particles emerging from, say, a scattering process.  Here
the flux at a detector is interpreted as the rate of arrival of the
scattered particles.

The additional assumption made in the BI is that particles exist
with simultaneously well defined positions and momenta and each
particle follows one of the one-parameter curves.  Such an
assumption is clearly excluded in standard quantum mechanics, but
this leaves us with the difficulty of understanding how to
incorporate the Born probability postulate and its role in the
continuity equation~(\ref{eq:sec2e10}) except in some abstract
sense.

If we do follow the BI in the $x$-representation, then the position
of the particle is clearly defined and the momentum, $p_{r}$,
associated with the particle must be provided through the relation
\begin{equation}
p_{r} = \frac{\Re \left[\psi^{*}(x)P\psi(x)\right]}{|\psi(x)|^{2}}
= \left(\frac{\partial S_{x}}{\partial x}\right)
= m\frac{j_{x}}{{\mathcal{P}}(x)} = m\frac{dx}{dt} \nonumber
\end{equation}
Here the ``beable'' momentum $p_{r}$ is wholly quantum in origin
showing that the BI has its origins entirely within quantum
mechanics.

Now in the $p$-representation we use
\begin{equation}
\frac{dp}{dt} = \frac{j_{p}}{{\mathcal{P}}(p)}
\end{equation}
to give a set of one-parameter curves in momentum space.  In this
approach the momentum of the particle has a clear meaning, while
the position beable $x_{r}$ is given by equation~(\ref{eq:n17}),
namely
\begin{equation}
x_{r} = \frac{\Re\left[\psi^{*}(p)X\psi(p)\right]}{|\psi(p)|^{2}}
= - \left(\frac{\partial S_{p}}{\partial p}\right)
\end{equation}
This means that the derivative in the current ${\bf{j}}_{p} = -
\Db{{\bf{p}}}\nabla_{X}(\rho
V({\bf{X}}))\Dk{{\bf{p}}}$ given by equation~(\ref{eq:n28}) must be
evaluated at $x = x_{r}$. Thus we again have a specification of the
particle with a given momentum at a given ``beable'' position $x_{r}$.

Thus, the BI based on equations~(\ref{eq:n15}) and~(\ref{eq:n16})
leads to two distinct phase spaces, one constructed on each
representation. Each phase space contains a set of trajectories,
one derived from $j_{x}$ and the other from $j_{p}$.  Although
these phase spaces are actually different, they carry structures
that are consistent with the content of the Schr\"{o}dinger
equation.  This is in contrast to the classical limit where there
is a unique phase space.  However we have already noted that
equations~(\ref{eq:n15}) and~(\ref{eq:n16}) reduce to a single
equation in the classical limit so the existence of (at least) two
phase spaces is a consequence of the quantum formalism.

Now the existence of at least two phase spaces may come as a
surprise to those who see the BI as a return to classical or
quasi-classical notions.  What we have shown here is that the BI
enables us to construct what we may call ``shadow phase spaces'', a
construct that is a direct consequence of the non-commutative
nature of the quantum algebra. Giving ontological meaning to the
non-commutative algebra implies a very radical departure from the
way we think about quantum processes.  This was the central theme
of Bohm's work on the implicate order \cite{DB80}.  The work
presented in this paper fits directly into this conceptual
structure, a point that will be discussed at length elsewhere.

Our present purpose is to clarify the structure of the mathematics
lying behind the BI.  To this end note that choosing a
representation is  equivalent to choosing an operator which is to
be diagonal.  Thus in the phase space described by
equation~(\ref{eq:n15}) the position eigenvalues are used for the
$x$ co-ordinates and we then construct the momentum co-ordinate
through the condition $p_{r} = (\partial S_{x}/\partial x)$  to
provide  the ``beable'' momentum.

On the other hand, equation~(\ref{eq:n16}) describes a phase space
constructed using the momentum eigenvalues together with the
``beable'' position $x_{r}$ defined by $x_{r} = - (\partial
S_{p}/\partial p)$. In this way we see exactly how it is possible
to  construct two {\it different} phase spaces, one for each
representation.

 The fact that we can find a BI in the $p$-representation removes
the criticism that the BI does not use the full symplectic symmetry
of the quantum formalism. But removing this asymmetry might, at
first sight, destroy the claim that the Bohm interpretation
provides a unique ontological interpretation.  This would only be
true if we were insisting that the ontology demands a unique phase
space. However, as we have already remarked the quantum algebra is
non-commutative and a unique phase space is not possible.  It was
for this reason that Bohm and one of us (BJH) began to explore the
possibility of giving ontological significance to the algebra
itself. This involves thinking in terms of process rather than
particles- or fields-in-interaction and this leads, in turn, to
introducing the implicate order mentioned above. This is a very
different order from the one assumed by most physicists, which is
essentially what we call the Cartesian order.

Once again we contrast our approach with Bohmian mechanics
introduced by D\"{u}rr et al \cite{DG}.  Their approach requires
the $x$-representation  to be taken as basic and the guidance
relation to be taken as {\it the} defining equation of the
approach. In view of the results presented here, we see we could
have started from the $p$-representation.  But here the relation $p
= (\partial S/\partial x)$ cannot play the role of a guidance
condition. Hence making the guidance condition as {\it the}
defining equation in the $x$-representation is arbitrary and
contrary to what Bohm himself had in mind \cite{DB50,DB57,BHDP}.

In regard to the lack of  $x$-$p$ symmetry in the traditional
approach to the BI, Bohm and Hiley \cite{BH93} found it necessary
to discuss why $x$ was the only intrinsic property of the particle,
all others depended upon the context. This was certainly felt by
one of us (BJH) to be a somewhat arbitrary imposition that did not
seem to be a natural consequence of the symplectic invariance of
the formalism itself. Had we started with the $p$-representation we
would have found $p$ to be the intrinsic property, while $x$
depended upon some context. Thus  the restoration of symmetry
explains why particular variables become intrinsic and others not.

In the examples we give in this paper, we only consider the two
operators $X$ and $P$. If  we  regard the change from the
$x$-representation to the $p$-representation as a rotation of
$\pi/2$ in phase space, we could think about exploring  rotations
through other angles.  Such transformation exist and are known as
fractional Fourier transformations which correspond to rotations
through any angle $\alpha$  in phase space \cite{AF1,AF2}. These
allow us to express equations~(\ref{eq:sec2e6})
and~(\ref{eq:sec2e7}) any arbitrary representation.  This
generalisation has been investigated and will be reported elsewhere
\cite{MB1}.

All of this shows that the $x$ variable is not special as far as
the mathematics goes.  The real question is why it is necessary to
construct different phase spaces in the first place, but before we
go into this question we want to present some examples where we can
compare in more detail the results obtained from both $x$- and
$p$-representations.
\newpage
\section{Specific examples: comparisons of $x$- and \\ $p$-representations}
\label{sec:spex}

\subsection{The free particle described by a Gaussian wave packet}

We will start with the simplest case of a particle described by a Gaussian
wave packet centred at position $x=0$ with mean momentum zero.

The wave packet has the (normalised) Gaussian distribution
\begin{equation}
\phi(p, t) = \left[\frac{2(\Delta x)^{2}}{\pi}\right]^\frac{1}{4}
\exp\left[- p^{2}(\Delta x)^{2}\right]\exp\left[-
\frac{ip^{2}}{2m}t\right]
\end{equation}
in the momentum representation.  In this
representation the current $j_{p} = 0 = (dp/dt)$ so that
the trajectories are of constant momentum.  Equation~(\ref{eq:sec2e9}) gives
\begin{equation}
\frac{\partial S}{\partial t} +\frac{p^{2}}{2m} = 0
\end{equation}
which shows that the quantum potential is zero, as is to be expected
from the form of the wave function $\phi (p, t)$.

In the $x$-representation, the wave packet spreads in the
$x$-direction, having the wave function
\begin{eqnarray}
&&\psi(x, t) = \frac{1}{(2{\pi}(D(t))^{\frac{1}{4}}}
\exp \nonumber \\
&&{\left[ -\frac{x^{2}}{4D(t)} + i\left(\frac{x^{2}t}{8m(\Delta
x)^{2}D(t)}-\frac{1}{2} \arctan {\left (\frac{t}{2m(\Delta x)^{2}}
\right)} \right) \right]}
\end{eqnarray}
where $D(t) = (\Delta x)^{2} + \left(\frac{t^{2}}{4m^{2}(\Delta
x)^{2}}\right)$.
The corresponding current is
\begin{equation}
j_{x} = \frac{{\mathcal{P}}(x)}{m}\frac{xt}{4m(\Delta x)^{2}D(t)}
\end{equation}
and equation~(\ref{eq:sec2e9}) yields
\begin{equation}
\frac{\partial S}{\partial t} + \frac{1}{2m}{\left(\frac{\partial S}
{\partial x}\right)}^2 +
\frac{1}{4mD(t)} - \frac{x^{2}}{8m[D(t)]^{2}} = 0,
\end{equation}
where the last two terms constitute the quantum potential.

This result can be easily understood since we are starting with the
particle confined in a region $\Delta x$ and, as time progresses,
the wave packet spreads out as expected.  The current $j_{x} =
{\mathcal{P}}(x)(dx/dt)$ and the trajectories calculated from this
current fan out in a way that exactly reflects the spread of
the wave packet. As the wave packet spreads, the quantum
potential reduces eventually to zero.  Thus, for a particular trajectory,
the energy of the quantum potential is progressively converted to the kinetic
energy of the particle, so accelerating it away from its
initial position.

\subsection{The quadratic potential}

Here we will simply collect the results derived earlier in the
paper for ease of comparison.

In the $x$-representation, where we write $\psi (x, t) = R_{x}
\exp[iS_{x}]$, the energy equation becomes
\begin{equation}
\frac{\partial S_{x}}{\partial t}+\frac{1}{2m}
\left(\frac{\partial S_{x}}{\partial
x}\right)^{2}+\frac{K}{2}x^{2}-\frac{1}{2mR_{x}}
\left(\frac{\partial ^{2}R_{x}}{\partial
x^{2}}\right)=0.
\label{eq:n29}
\end{equation}
While in  the $p$-representation, where we now write $\psi (p, t)
= R_{p}\exp[iS_{p}]$, the conservation of energy equation is
\begin{equation}
\frac{\partial S_{p}}{\partial t}+\frac{p^{2}}{2m}+
\frac{K}{2}\left(\frac{\partial
S_{p}}{\partial p}\right)^{2}
-\frac{K}{2R_{p}}\left(\frac{\partial ^{2}R_{p}}{\partial
p^{2}}\right)=0
\label{eq:n30}
\end{equation}

Now we turn to the probability currents and find
\begin{equation}
j_{x} = \frac{1}{2mi}\left[\psi ^{*}(x)
\left(\frac{\partial \psi(x)}{\partial x}\right) -
\left(\frac{\partial \psi^{*}(x)}{\partial x}\right)\psi(x)\right]
\end{equation}
\begin{equation}
j_{p} = \frac{K}{2i}\left[\psi^{*}(p)
\left(\frac{\partial \psi(p)}{\partial p}\right) -
\left(\frac{\partial \psi^{*}(p)}{\partial p}\right)\psi
(p)\right],
\end{equation}
in which the symmetry of the Hamiltonian is evident. That these
currents are in fact different should not be too surprising as they
arise in different spaces. Indeed, we can bring this out more
clearly by using the respective polar forms of the x- and
p-representation wave functions.  In the x-representation
\begin{equation}
j_{x} = \frac{1}{m}R_{x}^{2}
\left(\frac{\partial S_{x}}{\partial x}\right)
\end{equation}
and so
\begin{equation}
\frac{dx}{dt} = \frac{1}{m}\left(\frac{\partial S_{x}}{\partial
x}\right) = \frac{p_{r}}{m},
\end{equation}
whereas in the p-representation
\begin{equation}
j_{p} = KR_{p}^{2}
\left(\frac{\partial S_{p}}{\partial p}\right)
\end{equation}
so that
\begin{equation}
\frac{dp}{dt} = K\left(\frac{\partial S_{p}}{\partial p}\right) =
-\left(\frac{\partial V}{\partial x}\right)_{x = x_{r}}.
\end{equation}
Thus, we see that the currents provide the mathematical means of
constructing trajectories in the $x$-space and $p$-space
respectively.  It is a feature of both the linear potential and the
quantum harmonic oscillator that $\frac{dp}{dt} =
-\left(\frac{\partial V}{\partial x}\right)_{x = x_{r}}$, though
this is not generally true.

\subsection{The linear potential}

Here the potential is $V(x) = ax$. In this case the current
operators are
\begin{equation}
J_{x} = \frac{1}{2m}(\rho P + P \rho)
\end{equation}
and
\begin{equation}
J_{p} = - a\rho.
\end{equation}

In the $p$-representation we find
\begin{equation}
j_{p}  = \Db{p} J_{p} \Dk{p}  =  - {\mathcal{P}}(p) a
=  {\mathcal{P}}(p)\frac{dp}{dt}.
\end{equation}
This result is identical to that obtained from classical mechanics
through the equation
\begin{equation}
\frac{dp}{dt}  =  - \frac{\partial V}{\partial x} = - a
\end{equation}
and suggests that the p-representation trajectories lie on the
corresponding classical manifold.  Indeed,
equation~(\ref{eq:sec2e9}) gives
\begin{equation}
\frac{\partial S_{p}}{\partial t} + \frac{p^{2}}{2m} -
a\frac{\partial S_{p}}{\partial p} = 0.
\end{equation}
Now using $x_{r} = - (\partial S_{p}/\partial p)$, we find that the
corresponding energy equation is
\begin{equation}
\frac{\partial S_{p}}{\partial t} + \frac{p^{2}}{2m} + ax_{r} = 0,
\end{equation}
which has the same form as the classical Hamilton-Jacobi equation
with $x_{r} = x$.  This confirms that, for the p-representation of
the linear potential, there is no quantum potential and that the
trajectories are indeed classical.

We now compare these results with those for the $x$-representation.
Here the corresponding Schr\"{o}dinger equation is
\begin{equation}
\frac{d^{2}\psi}{dx^{2}} - A^{3}x\psi = 0
\end{equation}
where $A = {(2ma)}^{\frac{1}{3}}$. This equation has an Airy
function
\begin{equation}
\psi(x) = {C}Ai(Ax)
\end{equation}
as a solution, which, being real, implies a zero probability
density current
\begin{equation}
j_{x}  =  \Db{x} J_{x} \Dk{x} =
\frac{{\mathcal{P}}(x)}{m}\frac{\partial S_{x}}{\partial x} = {\mathcal{P}}(x)
\frac{dx}{dt} = 0.
\end{equation}
Using this result in equation~(\ref{eq:sec2e9}) shows that in the
$x$-representation the quantum potential is the negative of the
classical potential and is not zero as in the $p$-representation.
This example demonstrates that, while they are consistent with the
Schr\"{o}dinger equation, Bohm trajectories may be representation
dependent.

By way of explanation of the latter point, we observe that in the
$p$-representation the wave function is complex, its incoming and
outgoing components being separate on respectively the positive and
the negative $p$-domains.  On the other hand, in the
$x$-representation, the incoming and outgoing waves combine to
produce a real wave function.  In particular, the
$x$-representation solution may be split into incident and
reflected components using the relation
\begin{eqnarray}
&&Ai(Ax)+{\exp(-\frac{2}{3}i\pi)}{Ai({Ax}\exp(-\frac{2}{3}i\pi))}
\nonumber \\
&& + {\exp(\frac{2}{3}i\pi)}{Ai({Ax}\exp(\frac{2}{3}i\pi))} = 0.
\end{eqnarray}
Taking the incident and reflected wave function separately, one
obtains non-zero probability density currents and a non-zero
quantum potential.  The resulting trajectories are classical at
infinity but are non-classical near the origin, where reflection
takes place with an instantaneous change of sign in velocity.  This
is in contrast to the classical trajectory which turns smoothly at
the origin.  It is important to note that the trajectories of the
incident and reflected waves respectively do not embody the effects
of interference.  It is this interference, absent in the
$p$-representation, which produces a stationary trajectory for the
combined solution in the $x$-representation.

\subsection{The cubic potential}

The quantum phase equation~(\ref{eq:sec2e9}) in the
$x$-representation using $p_{r} = (\partial S_{x}/\partial x)$
gives
\begin{equation}
\frac{\partial S_{x}}{\partial t}+\frac{p_{r}^{2}}{2m} + Ax^{3} -
\frac{1}{2mR_{x}}\frac{\partial^{2}R_{x}}{\partial x^{2}} = 0,
\end{equation}
while in the $p$-representation
\begin{eqnarray}
&&\frac{\partial S_{p}}{\partial t}+\frac{p^{2}}{2m} +
A{x_{r}}^{3}+\frac{3A}{R_{p}}\frac{\partial^{2}R_{p}}{\partial
p^{2}}
\left(\frac{\partial S_{p}}{\partial
p}\right) \nonumber  \cr \cr && +
\frac{3A}{R_{p}}\left(\frac{\partial R_{p}}{\partial p}\right)
\left(\frac{\partial
^{2}S_{p}}{\partial p^{2}}\right)+
A\left(\frac{\partial ^{3}S_{p}}{\partial p^{3}}\right) = 0,
\end{eqnarray}
where we have used $x_{r} = - (\partial S_{p}/\partial p)$. This
clearly gives a far more complicated quantum potential.
Nevertheless, the content is still consistent with
Schr\"{o}dinger's equation. Both equations reduce to the same
classical Hamilton-Jacobi equation when the quantum potential terms
reduce to zero.

The respective currents are
\begin{equation}
j_{x} = \frac{1}{2mi}\left[\psi^{*}(x)
\left(\frac{\partial \psi(x)}{\partial x}\right) -
\left(\frac{\partial \psi^{*}(x)}{\partial x}\right)\psi(x)\right]
\end{equation}
and
\begin{equation}
j_{p} =
\frac{A}{2i}\left[\psi(p)\frac{\partial ^{2}\psi^{*}(p)}{\partial p^{2}} +
\psi^{*}(p)\frac{\partial ^{2}\psi(p)}{\partial p^{2}}
- \frac{\partial \psi(p)}{\partial
p}\frac{\partial \psi^{*}(p)}{\partial p}\right]
\end{equation}
which gives
\begin{equation}
j_{p} = - R_{p}^{2}\left(\frac{\partial V}{\partial x}\right)_{x = x_{r}}^{2}
+  A\left[2R_{p}\left(\frac{\partial
^{2}R_{p}}{\partial p^{2}}\right)
- \left(\frac{\partial R_{p}}{\partial p}\right)^{2}\right]
\end{equation}
as opposed to the simple expression for $j_{x}$
\begin{equation}
j_{x} = \frac{1}{m}R_{x}^{2}\left(\frac{\partial S_{x}}{\partial
x}\right).
\end{equation}
This clearly shows the limitation of using the condition $p_{r} =
(\partial S_{x}/\partial x)$ as the guidance condition.  It should
also by now be quite clear that the Bohm trajectories in a
particular representation are obtained from the probability current
for that particular representation and not from any additional
guidance condition.

\section{Conclusions}
\label{sec:conc}

\subsection{Algebraic formulation of the Schr\"{o}dinger picture}

In this paper we have shown how it is possible to write the content
of the Schr\"{o}dinger equation in algebraic form without reference
to either Hilbert space or to any specific representation. The
resulting two equations are respectively the Liouville equation,
equation~(\ref{eq:sec2e6}), and an equation that describes the time
development of the phase, equation~(\ref{eq:sec2e7}), which we have
called the quantum phase equation. Furthermore, we have shown that
this equation is gauge invariant and from it we calculated the
Aharonov-Bohm, the Aharonov-Casher and the Berry phases in a simple
and straight forward way.

We have also shown that it is possible to write the probability
currents as algebraic operator forms. This allows us to define
probability currents in any arbitrary representation.  All of these
results follow from the quantum formalism without the need to
appeal to any classical formalism.

\subsection{The $x$ and $p$ representations: the
quantum potential and the trajectories of probability current}

In sections~\ref{sec:tqpe} and~\ref{sec:pc}, we expressed
equations~(\ref{eq:sec2e8}) and~(\ref{eq:sec2e9}) in the
$x$-representation (equations~(\ref{eq:n25}) and~(\ref{eq:n15})
respectively) and showed that they are identical to the two
defining equations of the traditional Bohm interpretation
\cite{BH93}. The quantum potential emerges from
equation~(\ref{eq:n12}), which in turn comes directly from
equation~(\ref{eq:sec2e9}), showing that it cannot be ``dismissed
as artificial and obscuring the essential meaning of the Bohm
approach'' \cite{SG98} without missing some of the essential novel
features of quantum processes.

In particular, {\it observed} characteristic quantum phase or gauge
effects come directly from equation~(\ref{eq:sec2e9}).  As
Philippidis, Bohm and Kaye \cite {PBK} have shown many years ago,
the presence of the AB effect alters the quantum potential, which
in turn accounts for the fringe shifts. Furthermore, it is the
presence of the quantum potential that offers an explanation of
Einstein-Podolsky-Rosen-type correlations \cite{BH75}, as well as
quantum state teleportation \cite{MH99}.

In section~\ref{sec:reba}, we also showed that we can construct a
BI in the $p$-representation. Comparing representations shows very
clearly that the Bohm trajectories are simply the trajectories
associated with the probability currents of the standard theory.
The only assumption added to the standard quantum theory in the
Bohm-Hiley \cite{BH93} version of the BI is that particles have
simultaneously well defined positions and momenta and actually
follow these trajectories. Further, we claim that this position is
implicit in pragmatic quantum mechanics in which the probability
currents are assumed to be  related to particle fluxes.

\subsection{Shadow phase spaces}

The central point that emerges from our approach is that we can
construct two different phase spaces.  In the example of the
harmonic oscillator, the $x$-phase space is based on
equations~(\ref{eq:n14}) and~(\ref{eq:n15}), while the $p$-phase
space is built using equations~(\ref{eq:n16}) and~(\ref{eq:n17}).
The explication of different phase spaces was further exemplified
in section~\ref{sec:spex}.  The reason why we must resort to
constructing different phase spaces is not too difficult to see
once it is realised that we are dealing with a non-commutative
structure\footnote{ A simple example of this kind of structure will
be found in Hiley \cite{BH90}, and Hiley and Monk \cite{HM}.}.

For a commutative algebra, the Gel'fand construction allows us to
start from the algebra and re-construct the underlying manifold
\cite{JM}.  Here the points, the topology and the metric structure
of the manifold are all carried by the algebra.  No such
construction is possible for a non-commutative algebra. Thus, in
our case there is no underlying phase space with points that can be
specified by the pair of {\it observables} $(x, p)$. This is just
what the uncertainty principle is telling us. This is the
physicist's way of explanation why there can be no single, unique,
underlying continuous phase space.

Any attempt to produce a single phase space, such as is done in the
Wigner-Moyal approach, must necessarily contain unacceptable
features \cite{RS}. In this case, the probability distribution can
be negative in certain situations.  For these reasons, we must
follow what is usually done in non-commutative geometry and
construct shadow manifolds.

In this context, equation~(\ref{eq:n12}) provides an explanation as
to why the energy can be conserved when we attribute to the
particle at position $x$, the beable momentum $p_{r} =
\partial S_{x}/\partial x$.  Since it is a {\it constructed}
momentum and not a {\it measured} momentum, the kinetic energy will
not have the value necessary to conserve the total energy.  Thus we
need another term to ``carry'' this difference.  Since
equation~(\ref{eq:n15}) is the expression for the conservation of
energy, the last term on the RHS of equation~(\ref{eq:n15}) is the
place to ``store'' this energy difference.  This shows that the
quantum potential energy is an internal energy, and clearly does
not have an external source.

\subsection{Implications for the Bohm Interpretation of quantum mechanics}

Finally, we will briefly comment on the implications of the above
analysis on the BI. The traditional BI assumes the
$x$-representation is special, but the reasons for this were never
made clear. It was generally assumed that all physics must take
place in an {\it a priori} given space-time arena, a point of view
that we have called the Cartesian order.  Hence the attempt to use
the guidance condition as a defining equation for Bohmian mechanics
\cite{DG}. However our mathematical analysis above shows that this
condition is a contingent feature, which is dependent on the
asymmetry of the Hamiltonian.

On the other hand, if we take the quantum formalism as primary then
we must place our emphasis on the non-commutative structure of the
algebra of formalism. If we do this then attempts to focus on a
single phase space, which is equivalent to giving primary relevance
to space-time, will fail.  This in turn calls into question the way
we think about quantum processes.  Indeed Bohm has already argued
that we must abandon the Cartesian order and replace it by a
radically new approach to quantum phenomena which he called the
implicate order \cite{DB80}.  Here the ontology is provided by the
concept of {\it process} which is to be described by the
non-commutative algebra.  This is not a process in space-time, but
a process from which space-time is to be abstracted. Abstraction
here means to `make manifest' and the order that is made manifest
is called the explicate order.

The key point about this view is that there may be more than one
explicate order and that these explicate orders cannot be made
manifest together at the same time.  This can be regarded as a
direct consequence of the participatory nature of the quantum
process.  Thus the implicate order contains an ontological
complementarity, which is a necessary consequence of the
non-commutative structure.  In this picture the BI discussed above
is said to contain two explicate orders, one depending on the
$x$-representation and the other on the $p$-representation. These
are the shadow phase spaces.  Both are equally valid descriptions
of the outward appearance of a quantum process within a given
context.

Since our classical world is dominated by appearances in
space-time, we would expect the most relevant explicate order to be
that based on the $x$-representation, with the context being
provided by the classical world. This is the world in which we
place our apparatus and where our measurements take place. But
clearly we need to explore these ideas further as a number of
questions remain unanswered.  We will leave this discussion for
another paper.

\end{document}